\documentclass[10pt,aps,prb,twocolumn,amsmath,notitlepage,floatfix,footinbib,superscriptaddress,showpacs,showkeys]{revtex4-1}

\usepackage{graphicx}
\usepackage{hyperref}

\usepackage{colortbl}
\newcommand{\up}{\ensuremath{\uparrow}}
\newcommand{\down}{\ensuremath{\downarrow}}
\begin{document}

\title{Dynamics of Kondo voltage splitting after a quantum quench}

\author{Igor Krivenko}
\author{Joseph Kleinhenz}
\affiliation{Department of Physics, University of Michigan, Ann Arbor, Michigan 48109, USA}
\author{Guy Cohen}\email{gcohen@tau.ac.il}
\affiliation{School of Chemistry, Tel Aviv University, Tel Aviv 69978, Israel}
\author{Emanuel Gull}\email{egull@umich.edu}
\affiliation{Department of Physics, University of Michigan, Ann Arbor, Michigan 48109, USA}
\affiliation{Center for Computational Quantum Physics, Flatiron Institute, New York, New York 10010, USA}

\date{\today}

\begin{abstract}
We analyze the time-dependent formation of the spectral function of an Anderson impurity model in the Kondo regime within a numerically exact real-time quantum Monte Carlo framework.
At steady state, splitting of the Kondo peak occurs with nontrivial dependence on voltage and temperature, and with little effect on the location or intensity of high-energy features.
Examining the transient development of the Kondo peak after a quench from an initially uncorrelated state reveals a two-stage process where the initial formation of a single central Kondo peak is followed by splitting.
We analyze the time dependence of splitting in detail and demonstrate a strong dependence of its characteristic timescale on the voltage.
We expect both the steady state and the transient phenomenon to be experimentally observable.
\end{abstract}

\maketitle

\emph{\label{sec:intro}Introduction}. Interacting quantum many-body systems often exhibit highly entangled states that cannot be described within an independent particle formalism.
The Kondo effect in a quantum dot \cite{GoldhaberGordon1998, Cronenwett1998} coupled to noninteracting leads is the paradigmatic example for such a state, as the dot electrons hybridize with the leads to form a highly correlated Kondo singlet state \cite{HewsonBook}. This state  manifests itself as a sharp peak in the local density of states \cite{Cronenwett1998,Pustilnik2004}.
The establishment of Kondo correlations can be examined in a quantum quench scenario, where an initially uncorrelated state slowly develops a coherence peak over time \cite{Nordlander1999,Nghiem2017}.

In the presence of a voltage, the Kondo peak is strongly suppressed and splits into two smaller peaks \cite{DeFranceschi2002,Leturcq2005,Leturcq2006,Anders2008,Cohen2014}.
Previous work has argued that the peak-to-peak distance is given by the voltage \cite{Rosch2001,Fujii2003,Fritsch2010,Meir1993,Shah2006,Fugger2018} and that the split state is significantly less correlated than the equilibrium state \cite{Rosch2001}.
It is therefore natural to examine the establishment of splitting after a quench from an initially uncorrelated state, and to expect that this less correlated state forms on a timescale shorter than that of the equilibrium state.

Despite significant analytical progress \cite{Schoeller2009,Andergassen2011,Kennes2012,Wang2010,Lanata2012,Pletyukhov2012,Maslova2017}, an accurate investigation of this scenario requires numerical methods that are able to simulate the real-time evolution after a quench accurately, for times long enough to reach the steady state.
Additionally, a full account of the continuous lead spectrum is crucial for correct treatment of the nonequilibrium steady state.
The major families of numerical methods include the noncrossing approximation and its higher-order generalizations \cite{BeyondNCA}, wave-function-based methods \cite{MCTDH,Balzer2011,Gramsch2013,Lin2015,White2004,Daley2004}, real-time path integral techniques \cite{ISPI,Segal2010,Eckel2010,Weiss2013}, the time-dependent numerical renormalization group \cite{Anders2005,Anders2006,Nghiem2014,Nghiem2018,Schwarz2018}, hierarchical equations of motion \cite{Jin2008,Zheng2009,Wang2013,Wang2015}, the auxiliary master equation approach \cite{Schwarz2016,Dorda2014,Dorda2015,Dorda2017,chen2018auxiliary}, and a wide variety of quantum Monte Carlo methods \cite{Han2007,Dirks2010,Muehlbacher2008,RealTimeQMC,Schmidt2008,RealTimeCTAUX,Koga2013,NoneqCTINT,NoneqCTINT2019,Schiro2009,Muehlbacher2011,Dirks2013,Gull2011,Cohen2013,Cohen2014PRB,Antipov2016,Kubiczek2019}.
Most of these approaches fall short in at least one of the aforementioned requirements.
This situation has changed with the development of the numerically exact inchworm quantum Monte Carlo method \cite{Inchworm1,Inchworm2,InchwormCurrents,InchwormOpenSystems,Boag2018} that in many cases eliminates the dynamical sign problem and is thereby able to reach the relevant timescales.

In this Rapid Communication, we examine the voltage splitting of the Kondo peak in detail.
We focus on the time-dependent formation of the peak after a quantum quench and on its shape at long times.
We find that while the peak-to-peak distance is roughly proportional to the voltage, there is a notable deviation from this simple picture.
We also find that the appearance of the split peak is preceded by the formation of a single, unsplit Kondo peak, and that the splitting occurs at a later time whose scaling with the voltage is consistent with a power law.
Since the splitting timescale is ~1--10 ps in mesoscopic quantum dots, the delayed splitting should be observable in recently developed ultrafast tunneling microscopy \cite{Cocker2013,Loth2010} and spectroscopy \cite{Eisele2014,ochoa_pumpprobe_2015} experiments.

\emph{\label{sec:model}Model}. We describe a correlated quantum dot (QD) attached to two extended metallic leads using a single impurity Anderson model \cite{Anderson1961},
\begin{subequations}
    \begin{align}\label{eq:hamiltonian}
        \hat H = & \hat H_D + \sum_{\alpha=\pm1} \hat H_{\alpha} + \hat H_T,\\
        \hat H_D = & \sum_{\sigma}\varepsilon_{d} n_{\sigma} +
                   Un_{\uparrow}n_{\downarrow}, \\
        \hat H_\alpha = & \sum_{k \sigma}
        \left(\varepsilon_{k} + \frac{\alpha V}{2}\right) n_{\alpha k\sigma}, \\
        \hat H_T = & \sum_{\alpha k\sigma}\mathcal{V}_k^\alpha(t)
        (c_{\alpha k\sigma}^{\dagger}d_{\sigma}
        + d_{\sigma}^{\dagger}c_{\alpha k\sigma}).
    \end{align}
\end{subequations}
The quantum dot $\hat H_D$ is coupled to two noninteracting leads $\hat H_\alpha$ by tunneling terms $\hat H_T$.
The operators $d_\sigma^\dagger$ ($d_\sigma$) create (annihilate) electrons localized on the quantum dot, while $c_{\alpha k\sigma}^\dagger$ ($c_{\alpha k\sigma}$) create (annihilate) electrons in lead $\alpha$ [$\alpha = \pm 1$ labels the left ($+$) and right ($-$) lead] with quasimomentum $k$ and spin $\sigma$ ($\uparrow$ or $\downarrow$).
The respective occupation number operators are $n_\sigma = d^\dagger_\sigma d_\sigma$ and $n_{\alpha k\sigma}=c_{\alpha k\sigma}^\dagger c_{\alpha k\sigma}$.
The dot Hilbert space is spanned by four ``atomic states'' $|\phi\rangle = {|0\rangle}, {|\up\rangle}, {|\down\rangle}, {|\up\down\rangle}$.
We consider the symmetric situation $\varepsilon_d = -U/2$ such that every energy level of the dot Hamiltonian $\hat H_D$ is doubly degenerate ($E_0 = E_{\up\down} = 0$, $E_\up = E_\down = -U/2$).
$\mathcal{V}_k^\alpha$ denotes the tunneling matrix element describing hopping processes between the dot and the leads.
The coupling to the leads is characterized by a coupling density $\Gamma_\alpha(\omega) = \pi\sum_k |\mathcal{V}_k^\alpha|^2 \delta(\omega-\varepsilon_k)$ that parametrizes the lead dispersion $\varepsilon_k$ and the tunneling elements.
We consider a wide, flat coupling density with soft edges for both leads, $\Gamma_\alpha(\omega) = (\Gamma/2) / [(1+ e^{\nu(\omega-D)})(1+ e^{-\nu(\omega+D)})]$ (the soft edges eliminate unphysical transient oscillations in the dynamics \cite{RealTimeCTAUX}), choosing the inverse cutoff width $\nu = 10\Gamma^{-1}$ and the half-bandwidth $D = 10\Gamma$ such that the band edge exceeds all other relevant energy scales.
$\Gamma$ is used as the energy unit. Experimental values for $\Gamma$ in semiconductor QDs are of the order of 1\,meV \cite{GoldhaberGordon1998,Cronenwett1998}.
We consider a setup where the dot is initially empty (in the pure ${|0\rangle}$ state) and detached from the leads.
The leads are suddenly attached at $t=0$ ($\mathcal{V}_k^\alpha(t) = \mathcal{V}\theta(t)$), and are kept at a constant temperature $T$ with a symmetric bias voltage $V$ between them. This quench protocol is equivalent to suddenly changing a gate voltage from a value substantially larger than half the bias voltage to zero at $t=0$.
At zero bias, the Kondo temperature for this model is $T_K \simeq \sqrt{\Gamma U/2}\exp[-\pi U / (8\Gamma) + \pi\Gamma / (2U)]$ \cite{Wiegmann1983,HewsonBook}.

\emph{\label{sec:method}Methods}. The numerical methods we use in this Rapid Communication are based on a diagrammatic expansion in the tunneling Hamiltonian $\hat H_T$ formulated on the two-branch Keldysh contour (the imaginary Matsubara branch is not required due to the factorized initial condition).
Our main numerical tool is a massively parallel implementation of the inchworm quantum Monte Carlo solver \cite{Inchworm1,Inchworm2}  based on the High Performance ParalleX framework \cite{HPX} and the ALPS libraries \cite{ALPSCore2017, ALPSCore2018}.
The inchworm solver performs a stochastic summation of the hybridization contributions to the dressed QD propagators $p_\phi(t,t') = \langle\phi| \mathrm{Tr}_{c}[\hat{\rho}e^{-i\int_{t'}^{t}d\bar t \hat H(\bar t)}]|\phi\rangle$.
The calculations are organized to take advantage of the contour-causal structure of $p_\phi(t, t')$ so that short time propagators are incrementally extended to longer times, significantly alleviating the dynamical sign problem \cite{InchwormCurrents,Inchworm2}.
After all dressed propagators are computed, the stochastic summation procedure proposed in \onlinecite{InchwormCurrents} is employed to calculate the QD Green's function.
Because the inchworm method recurrently couples together the output of many stochastic simulations, the analysis of the Monte Carlo error is not straightforward.
One useful approach is considering deviations from exactly conserved properties like the total probability of all QD states or the normalization of the steady-state spectral function, neither of which varies by more than a few percent in our simulations.
Using the inchworm method we obtained numerically exact results for times as long as $8.0\Gamma^{-1}$, but required significant computational resources to do so.
In order to investigate longer times, we also make use of the computationally less demanding one crossing approximation (OCA) \cite{Pruschke1989,BeyondNCA} which we validate against numerically exact inchworm results at our smallest considered interaction strength $U = 8.0\Gamma$ where OCA is expected to be the least accurate.
We find that while details of the spectral function are rather sensitive to this approximation (see Supplemental Material for a comparison \cite{suppl}), it is accurate to within $\lesssim10\%$ for the other observables considered here.

The main physical quantity of interest to us is the time-dependent QD spectral function. We use the auxiliary current formalism \cite{Sun2001,Cohen2014PRB} to write this as
\begin{equation}
    A(\omega, t) = \lim_{\eta\to 0}-\frac{2h}{e\pi\eta} [I_A^f(\omega,t) - I_A^e(\omega,t)],
\end{equation}
where $I_A^f(\omega,t)$ and $I_A^e(\omega,t)$ are currents through two additional auxiliary leads weakly coupled to the QD at frequency $\omega$ by a coupling density $\Gamma_A(\omega') = \eta\delta(\omega' - \omega)$, and with chemical potentials set such that the leads are full and empty, respectively.
$A(\omega, t)$ approaches the conventional spectral function $A(\omega) = -(1/\pi)\mathrm{Im} G^r(\omega)$ at steady state and provides rich spectral information at all times. It is related to a finite-time Fourier transform, but also has a direct operational realization \cite{Lebanon2001,Sun2001,Cohen2014PRB,Cohen2014}.
We have direct access to the QD Green's function $G_{\sigma}(t,t') = -i\langle\mathit{T}_\mathcal{C} d_\sigma(t)d^\dagger_\sigma(t')\rangle$ \cite{InchwormCurrents}, such that auxiliary currents are calculated using the Meir-Wingreen formula \cite{Meir1992, QiaoyuanThesis} $I_A^{f(e)}(\omega,t) = -2 \mathrm{Re}\left\{\int_\mathcal{C} dt' G_\sigma(t',t) \Delta_A^{f(e)}(t, t') \right\}$. Here, the hybridization functions $\Delta_A^{f(e)}(t, t')$ are derived from $\Gamma_{A}(\omega)$ using the procedures established in Ref.~\onlinecite{InchwormCurrents}.

\begin{figure}
    \includegraphics[width=\columnwidth]{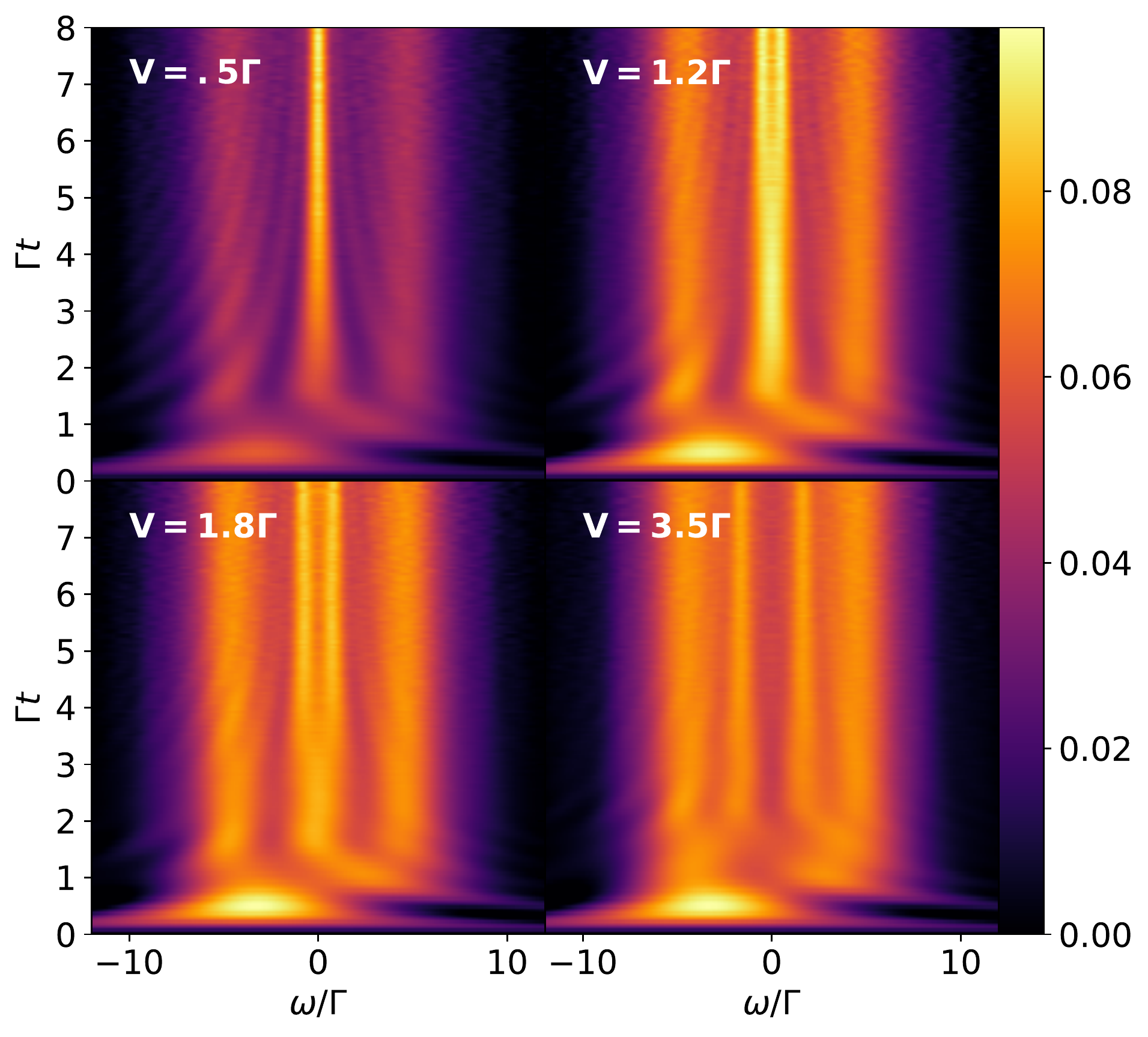}
    \caption{\label{fig:U8_beta50_aux_spectra}
        Time evolution of the QD spectral function after a coupling quench and in the presence of a bias voltage $V$, at interaction strength $U=8.0\Gamma$ and temperature $T=0.02\Gamma \ll T_K$. The voltages are $V=0.5\Gamma$ (upper left),
        $V=1.2\Gamma$ (upper right), $V=1.8\Gamma$ (lower left), and
        $V=3.5\Gamma$ (lower right).
    }
\end{figure}

\emph{\label{sec:results}Results}. In Fig.~\ref{fig:U8_beta50_aux_spectra}, we present the time evolution of the (auxiliary) QD spectral function after a coupling quench.
The time-dependent spectra are shown at four values of the bias voltage, $V = 0.5\Gamma, 1.2\Gamma, 1.8\Gamma$, and $3.5\Gamma$.
The interaction strength $U$ is $8.0\Gamma$, such that $T_K\approx0.11\Gamma$ \cite{HewsonBook}.
The lead temperature is set to $T=0.02\Gamma \ll T_K$, placing the system deep in the Kondo regime at zero bias. This temperature was inaccessible in the earlier bold-line hybridization expansion QMC study [\onlinecite{Cohen2014}], where only the edge of the Kondo regime $T\gtrsim T_K$ was reached at a weaker interaction strength $U=6\Gamma$.

For $V \lesssim 2.0\Gamma$, we observe the formation of a single peak at the mean chemical potential.
For $V \gtrsim 1.0\Gamma$, this peak first forms, then splits into two secondary peaks near the two lead chemical potentials.
At larger $V \gtrsim 2.0\Gamma$, the initial single-peak state is no longer visible.
Instead, the split peaks appear immediately after the transient charging dynamics visible at short times.

\begin{figure}
    \includegraphics[width=\columnwidth]{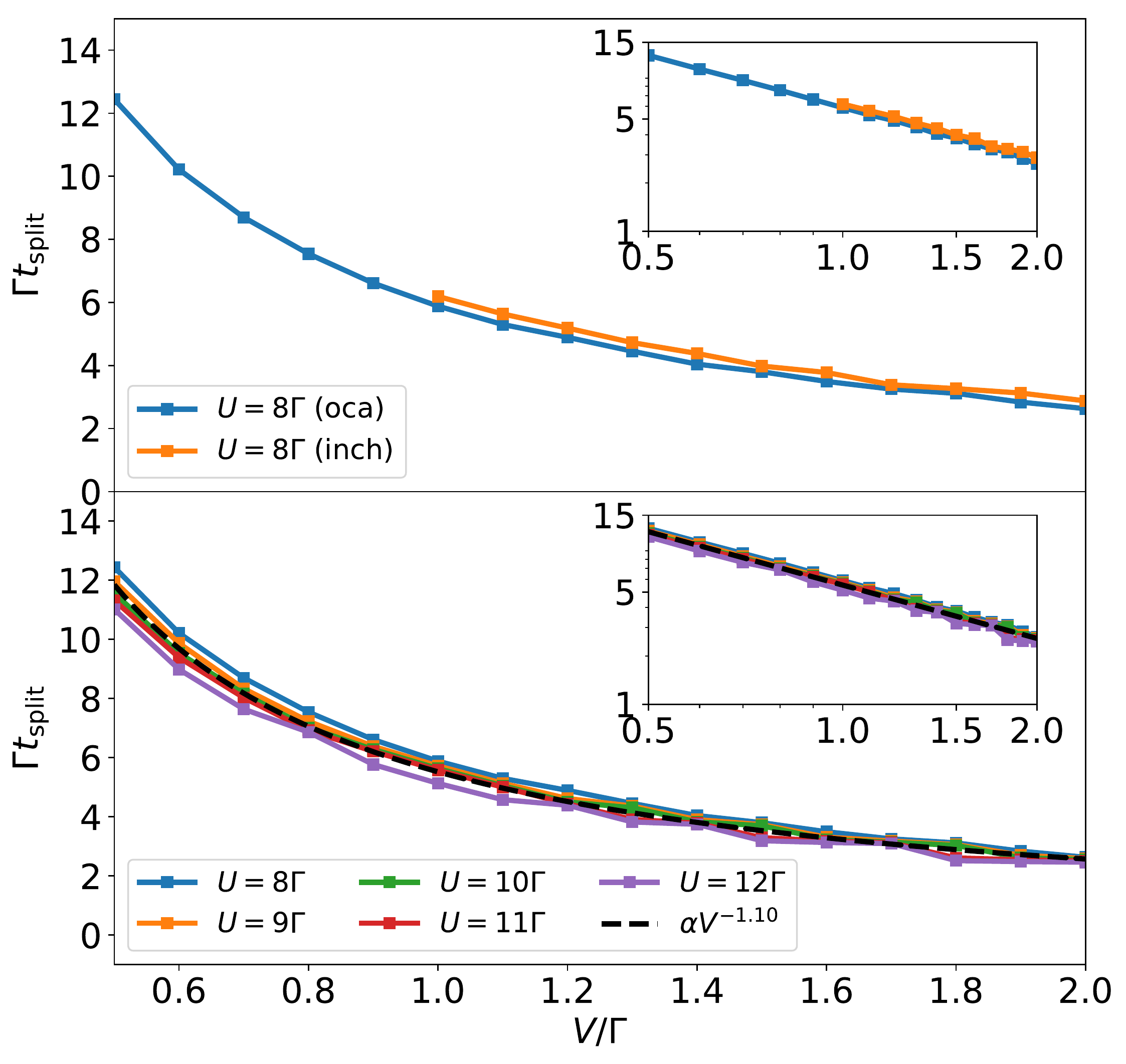}
    \caption{\label{fig:t_split}
        Splitting time $t_\text{split}$ where a single peak splits into two peaks as a function of $V$ at $T = 0.02\Gamma$.
        Upper panel: inchworm and OCA results at $U = 8.0\Gamma$.
        Lower panel: OCA results for several $U$.
        Insets show the same data on a log-log scale demonstrating power-law 
        behavior.
    }
\end{figure}

The overall behavior of the system can be characterized by two quantities: the time $t_\text{split}$ at which the peak splits, and the peak-to-peak separation $\Delta\omega$ of the resulting split peaks in steady state.
We first show how $t_\text{split}$ evolves as a function of the bias voltage $V$ at $T = 0.02\Gamma$.
$t_\text{split}$ is calculated as the first time point where the second derivative $\partial^2_\omega A(\omega, t)|_{\omega=0}$ changes its sign, i.e. where the zero-frequency peak becomes a dip.
As seen in the upper panel of Fig.~\ref{fig:t_split}, the OCA is qualitatively consistent with inchworm regarding the functional form of $t_\text{split}(V)$.
A log-log plot (inset) reveals that within the voltage range shown both results are consistent with power-law behavior, which from the slope must obey $t_\text{split}(V)\propto(V/\Gamma)^{-1.10}$.
OCA results for a set of larger interaction strengths (lower panel of Fig.~\ref{fig:t_split}) provide evidence that this behavior is largely independent of $U$.
It is worth noting that the transient state manifested by the single peak is significantly less correlated than the equilibrium Kondo singlet at $V=0$. One can indirectly assess the strength of correlations by fitting the transient spectra with the equilibrium ones obtained for an effective temperature $T_\text{eff}^\text{tr}$. Similarly, the steady-state spectra $A(\omega, V, T)$ can be qualitatively fitted with $[A(\omega+V/2, 0, T_\text{eff}^\text{st}) + A(\omega-V/2, 0, T_\text{eff}^\text{st})] / 2$ (superposition of contributions from two Kondo states with different chemical potentials and effective temperature $T_\text{eff}^\text{st}$). The relevant figures are Figs. S3 and S4 of Ref.~\onlinecite{suppl}. It turns out that the effective temperature, serving as a measure of correlations, satisfies $T_\text{eff}^\text{tr} \gg T_\text{eff}^\text{st} \gtrsim T_K \gg T$.

An analysis at $T>T_K$ shows that the initial single-peak state is not present for $T > T_K$.
Instead, the voltage-split peaks are formed directly after the initial equilibration (see Fig.~S1 in \onlinecite{suppl}).

Our results suggest that the time-dependent formation of the spectrum evolves in two stages.
First, on a very fast timescale, a mixed Kondo singlet is formed between the QD and an effective chemical potential set by those of both leads.
Later, on a slower timescale $t=t_\text{split}$, this singlet state is destroyed by the current and replaced with a new state that couples to each of the two leads at a frequency comparable to its chemical potential.

\begin{figure}
    \includegraphics[width=\columnwidth]{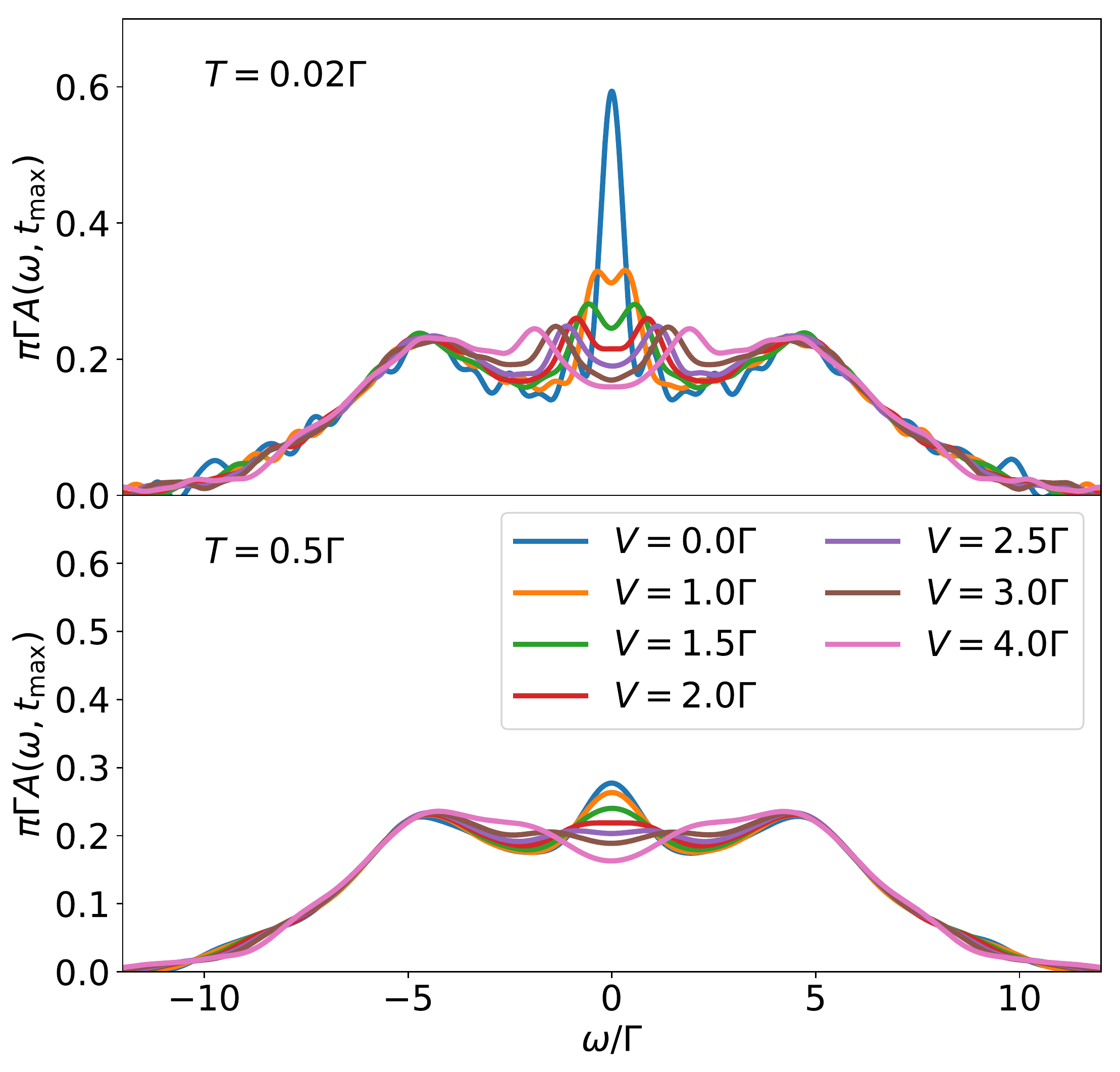}
    \caption{\label{fig:U8_aux_spectra}
        QD spectral function from the inchworm method at $t = t_\text{max} = 8.0\Gamma^{-1}$, corresponding to the steady-state spectrum $A(\omega)$ for $V\geq1.0\Gamma$.
        $T=0.02\Gamma$ (upper panel) and $T=0.5\Gamma$ (lower panel) for the voltages indicated.
    }
\end{figure}

So far, we have focused on the dynamics leading up to the formation of a steady state. We now shift to a discussion of the frequency-dependent spectral properties of the steady state itself. Figure~\ref{fig:U8_aux_spectra} provides a detailed view of $A(\omega,t_\text{max})$, which gives an estimate of the steady-state spectra for $V\geq 1.0\Gamma$ both below (upper panel) and above (lower panel) the Kondo temperature at $U=8.0\Gamma$.
At low temperature ($T=0.02\Gamma < T_K$) and intermediate bias voltage ($1.0\Gamma \leq V \leq 2.0\Gamma$) the split Kondo peaks and the Hubbard bands together form a clearly distinguishable four-peak structure, confirming previous approximate results that suggested its existence \cite{wingreen_anderson_1994,Dorda2014} (see also very recent results where partial splitting is visible in Ref.~\onlinecite{NoneqCTINT2019Results}).
An increase in $V$ enhances peak-to-peak separation $\Delta\omega$ and suppresses peak height \cite{Rosch2001}.
In contrast, the side bands are largely insensitive to changes in $V$ in this regime. (The small rapid oscillations seen at $V=0$ are remnants of the initial condition that have not fully dissipated.)
At high temperature $(T=0.5\Gamma > T_K)$, some remnants of the four-peak structure can be seen between $V=2.5\Gamma$ and $V=3.0\Gamma$. However, these features are much less pronounced.

\begin{figure}
    \includegraphics[width=\columnwidth]{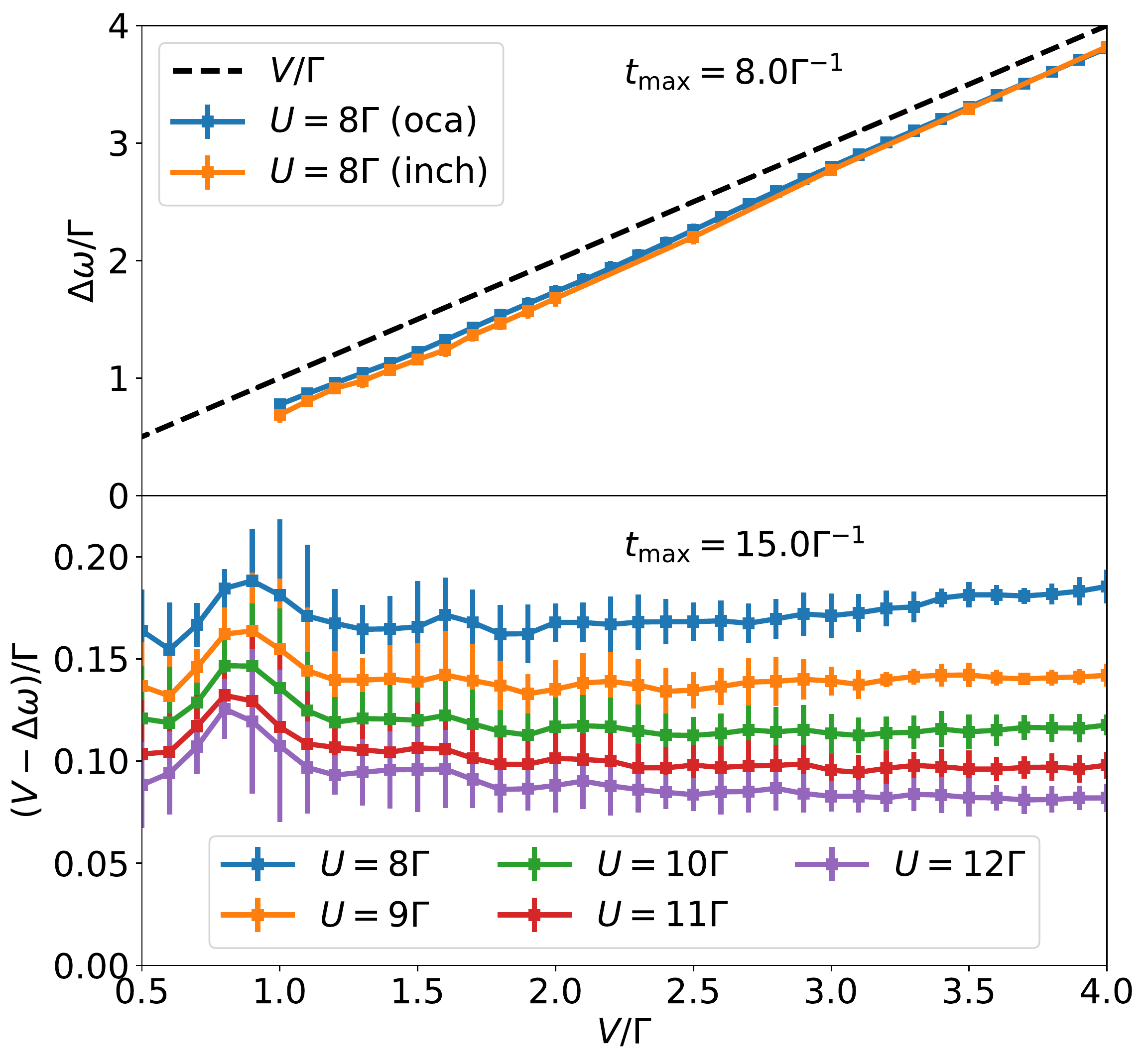}
    \caption{\label{fig:kondo_splitting}
        Peak-to-peak distance $\Delta\omega$ between the split Kondo peaks as a function of bias voltage $V$, at $T = 0.02\Gamma < T_K$. Error bars originate from averaging over finite-time oscillations expected to eventually dissipate.
        Upper panel: inchworm and OCA results at $U = 8.0\Gamma$ with $t_\text{max} = 8.0\Gamma^{-1}$ compared with the linear behavior $\Delta\omega = V$ predicted by various approximate methods.
        Lower panel: Deviation of splitting from $V$ within OCA at $T = 0.02\Gamma$ for several values $U$, with $t_\text{max} = 15.0\Gamma^{-1}$.}
\end{figure}

In Fig.~\ref{fig:kondo_splitting}, we present the peak-to-peak distance $\Delta\omega$ at steady state as a function of the applied bias voltage $V$.
Parameters match the respective panels in Fig.~\ref{fig:t_split}.
We estimate the steady-state value from propagation to a finite time. Although the large-scale features of the spectrum have reached steady state by our maximal propagation time, there remain small finite-time oscillations that are expected to eventually dissipate. The error bars in this figure therefore come from averaging $\Delta\omega$ over the time window in which splitting is visible.
As seen in the upper panel of Fig.~\ref{fig:kondo_splitting}, $\Delta\omega$ is systematically below the linear $\Delta\omega=V$ law predicted by perturbation theory, renormalization group, and flow equation studies of the Anderson model \cite{Fujii2003} and the effective {\it s-d} (Kondo) model \cite{Fritsch2010, Rosch2001}.
We reiterate that the inchworm results presented here are numerically exact, whereas the various approximate approaches are, e.g., perturbative in $U$ or assume a $U\to\infty$ limit where charge fluctuations on the QD are suppressed.
We expect our prediction to be experimentally verifiable using steady-state multiprobe schemes \cite{Lebanon2001,Sun2001}.

If we acknowledge that the trend evident in Fig.~\ref{fig:t_split} may continue to smaller bias voltages $V < 1.0\Gamma$, it is clear that no conclusion about the presence of splitting can be drawn from the inchworm results at lower voltages because $t_\text{split}$ can exceed $t_\text{max}$.
We employ OCA in order to reach longer times $t_\text{max} = 15.0\Gamma^{-1}$ and explore a wider parameter range. The larger $t_\text{max}$ in these OCA results extends the accessible voltage range down to $V = 0.5\Gamma$, but at the cost of introducing an approximation.
To test the quality of this approximation, the upper panel of Fig.~\ref{fig:kondo_splitting} shows numerically exact inchworm data together with OCA results at $t_\text{max} = 8.0\Gamma^{-1}$.
The agreement in $\Delta\omega/\Gamma$ between inchworm and OCA is on the order of 10\% and improves at larger $V$, with OCA somewhat underestimating the deviation from linear behavior at smaller $V$.
This observation suggests that electronic correlations beyond those accounted for by the OCA become less important as the bias voltage grows.
Together with the diminishing height of the split peaks, this supports the scenario in which the Kondo state is partially destroyed by the current-induced decoherence  \cite{Rosch2001}.

The lower panel of Fig.~\ref{fig:kondo_splitting} shows OCA results for $V - \Delta\omega$ at several larger values $U$ where OCA is expected to be increasingly accurate.
These OCA calculations show that $V - \Delta\omega$ becomes smaller with increasing $U$. This supports the conclusion that deviations from the linear approximation are due to charge fluctuations at finite $U$.

\emph{\label{sec:conclusions}Conclusions}. We presented a numerically exact treatment of the transient and steady-state dynamics of a quantum dot spectral function after a coupling quench with a bias voltage $V$ applied to the dot, focusing on the Kondo regime.

Our examination of the quench dynamics revealed transient dynamical states in which the formation of a single Kondo peak at the average chemical potential is followed by a sudden splitting at a timescale $t_\text{split}$. $t_\text{split}$ exhibits a robust power-law dependence on the voltage. In the case of realistic molecular electronic devices $\Gamma\approx 100$ meV. At a voltage of $\sim50$ meV our predicted timescale approaches $t_\text{split}\sim10^{-1}$ ps, but if the power law holds at lower voltages, at a voltage of $\sim5 $ meV we expect $t_\text{split}\sim1$ ps, which is already experimentally accessible. Furthermore, in semiconductor quantum dot experiments $\Gamma$ is orders of magnitude smaller, e.g. $0.1-1.0$ meV according to Refs.~\onlinecite{GoldhaberGordon1998} and \onlinecite{Cronenwett1998}. A typical high-voltage $t_\text{split} = 5\Gamma^{-1}$ would then correspond to $\sim3$--$30$ ps.
These predictions concern the transient dynamics of the time-dependent spectral density. Although measuring it is still challenging, recent experimental progress \cite{Cocker2013,Eisele2014,ochoa_pumpprobe_2015} may put it within reach. One possible direction is to extract the time dependent current from DC measurements with pulse trains as suggested by Ref.~\onlinecite{Eisele2014} in a three-terminal setup.

For voltages significantly exceeding the Kondo temperature, we presented numerically exact results for steady-state spectral functions exhibiting a well pronounced four-peak structure.
The position and shape of the side bands are unaffected by the bias voltage $V$, while the distance $\Delta\omega$ between the split Kondo peaks is roughly proportional to $V$ but systematically falls below the previously proposed $\Delta\omega = V$ behavior. This effect weakens at large $U$, and we therefore surmise that it is related to charge fluctuations that are energetically forbidden when $U$ becomes very large. These predictions could be verified using three-terminal steady-state measurements as discussed in Refs.~[\onlinecite{Sun2001}] and [\onlinecite{Lebanon2001}].

Our application of the inchworm method to exploring nonequilibrium Kondo physics after a quench elucidates the dynamical formation of Kondo splitting, and provides experimentally relevant predictions thereof. Looking forward, this work points the way towards answering a variety of long-standing questions, such as whether further splitting should be expected when a magnetic field is present; how correlations form in the leads; and how local symmetries affect the Kondo coupling far from equilibrium. Another interesting direction is application of the inchworm QMC to direct modeling of response to realistic short pump pulses, as used in pump--probe experiments.

\emph{\label{sec:acknowledgments}Acknowledgments.} The authors are grateful to Vladimir Mantsevich and Sergei Iskakov for helpful discussions.
This work was supported by DOE ER 46932. Our research used computing resources of the National Energy Research Scientific Computing Center (NERSC), a U.S. Department of Energy Office of Science User Facility operated under Contract No. DE-AC02-05CH11231.
G.C. acknowledges support by the Israel Science Foundation (Grant No. 1604/16).
International collaboration was supported by Grant No. 2016087 from the United States--Israel Binational Science Foundation (BSF).

\bibliographystyle{apsrev4-1}
\bibliography{kondo_voltage_splitting}

\end{document}